# The Gain in the Field of Two Electromagnetic Waves

K.V. Ivanyan*

M.V. Lomonosov Moscow State University, Moscow 119991, Russia

We consider the motion of a nonrelativistic electron in the field of two strong monochromatic light waves propagating counter to each other. The matrix elements of emission and absorption are found. An expression is obtained for the gain of a weak test wave by using such matrix elements.

## INTRODUCTION

The motion of an electron in the field of a monochrornatic light wave is described by the well-known volkov function. Exact solutions of the relativistic wave equations were obtained in [1,2] for the motion of an electron in certain cases of plane-wave fields. The motion of an electron in the field of two light waves propagating counter to each other (standing wave), however, cannot be solved exactly.

Electron diffraction by a standing wave was considered by perturbation theory in [3]. The electron channeling was investigated in in intense standing light wave in [4-6]. Other schemes for FEL were considered in [7-57].

Using the wave function derived in [58], we obtain the gain of a weak test wave. We show that the gain is an optimum if the angle $\theta$ between **p** and **k** is close to $\pi/2$.

## GENERAL RELATIONS

In [58] was found wave function of electron in the field of two strong monochromatic light waves propagating counter to each other formula (21).

We now define the field of the amplified electromagnetic wave by the relation

$$\mathbf{A}_3(\mathbf{r},t) = \mathbf{A}_3 \sin(\omega_3 t - \mathbf{k}_3 \mathbf{r}), \tag{1}$$

where $\omega_3$ is the frequency of the amplified wave and $\mathbf{k}_3$ is its wave vector. The processes investigated in this paper are charcterized by an element of an S matrix for which the expressions take, in first-order perturbation theory in the test-wave electromagnetic field, the form

-----------------

*k.ivanyan@yandex.com

$$S_{fi} = -i\int \Psi_f^*(\mathbf{r},t)\hat{V}\Psi_i(\mathbf{r},t)d\mathbf{r}dt, \tag{2}$$

$$\hat{V} = e\mathbf{A}_3(\mathbf{r},t)p/mc. \tag{3}$$

The S-matrix element is, with allowance for relation (3),

$$\begin{aligned}
S_{fi} &= \frac{e\mathbf{A}_3\mathbf{p}}{2mcV}\int \exp\left[i\left(\frac{E-E'}{\hbar}\pm\omega_3\right)t + i\left(\frac{p-p'}{\hbar}\pm k_3\right)r\right] \\
&\times \exp\left[i\frac{e\mathbf{A}_1(\mathbf{p}'\text{-}\mathbf{p})}{mc\hbar\omega_1}\cos(\omega_1 t - \mathbf{k}_1\mathbf{r}) + i\frac{e\mathbf{A}_2(\mathbf{p}'\text{-}\mathbf{p})}{mc\hbar\omega_2}\cos(\omega_2 t - \mathbf{k}_2\mathbf{r})\right] \\
&\times \frac{e\mathbf{A}_1\mathbf{A}_2(\mathbf{v}\text{-}\mathbf{v}')(\mathbf{k}_2\text{-}\mathbf{k}_1)}{2mc^2\hbar\{[\omega_1 - \omega_2 + \mathbf{v}(\mathbf{k}_2\text{-}\mathbf{k}_1)]\}^2} d\mathbf{r}dt
\end{aligned} \tag{4}$$

(the + and - signs pertain to emission and absorption, respectively) and where $\mathbf{A}_1(\varphi_1)$ and $\mathbf{A}_2(\varphi_2)$ are the vector potentials of the electromagnetic waves, respectively; where $\varphi_1 = \omega_1 t - \mathbf{k}_1\mathbf{r}$, $\varphi_2 = \omega_2 t - \mathbf{k}_2\mathbf{r}$ ($\omega_1, \omega_2$ are respectively the frequencies of the first and second electromagnetic waves, and $\mathbf{k}_1$ and $\mathbf{k}_2$ are the wave vectors of these waves); $m$ is the electron mass.

We use here the known relation

$$\exp[iB\sin x] = \sum_{n=-\infty}^{+\infty} J_n(B)e^{inx}. \tag{5}$$

The expression for the S-matrix element is then

$$\begin{aligned}
S_{fi} &= -i\frac{e\mathbf{A}_3\mathbf{p}}{2mcV}\int \sum_{n_1,n_2} J_{n_1}\left(\frac{e\mathbf{A}_1(\mathbf{p}'\text{-}\mathbf{p})}{mc\hbar\omega_1}\right)J_{n_1}\frac{e\mathbf{A}_2(\mathbf{p}'\text{-}\mathbf{p})}{mc\hbar\omega_2} \\
&\times \exp\left\{i\left[\frac{(\mathbf{p}\text{-}\mathbf{p}')}{\hbar}\mp\mathbf{k}_3 + n_1\mathbf{k}_1) + n_2\mathbf{k}_2 + (\mathbf{k}_2\text{-}\mathbf{k}_1)\right]\mathbf{r}\right\} \\
&\times \frac{e^2\mathbf{A}_1\mathbf{A}_2(\mathbf{p}\text{-}\mathbf{p}')(\mathbf{k}_2\text{-}\mathbf{k}_1)}{(2mc^2)^2\hbar\{\omega_1 - \omega_2 + \mathbf{v}(\mathbf{k}_2\text{-}\mathbf{k}_1)\}^2} \\
&\times \exp\left[i\left(\frac{E'-E}{\hbar}\pm\omega_3 + n_1\omega_1 + n_2\omega_2 + \omega_1 - \omega_2\right)t\right]d\mathbf{r}dt.
\end{aligned} \tag{6}$$

Integration with respect to r and t in (6) leads to the following expression for the S-matrix element:

$$\begin{aligned}
S_{fi} &= -i\frac{e\mathbf{A}_3\mathbf{p}(2\pi)^4}{2mcV}\sum_{n_1,n_2}J_{n_1}\left(\frac{e\mathbf{A}_1(\mathbf{p}'\text{-}\mathbf{p})}{mc\hbar\omega_1}\right)J_{n_1}\frac{e\mathbf{A}_2(\mathbf{p}'\text{-}\mathbf{p})}{mc\hbar\omega_2}\frac{e^2\mathbf{A}_1\mathbf{A}_2(\mathbf{p}\text{-}\mathbf{p}')(\mathbf{k}_2\text{-}\mathbf{k}_1)}{(2mc^2)^2\hbar\{\omega_1-\omega_2+\mathbf{v}(\mathbf{k}_2\text{-}\mathbf{k}_1)\}^2} \\
&\times \delta\left(\frac{E'-E}{\hbar}\pm\omega_3+n_1\omega_1+n_2\omega_2+\omega_1-\omega_2\right)\delta\left[\frac{(\mathbf{p}\text{-}\mathbf{p}')}{\hbar}\mp\mathbf{k}_3+n_1\mathbf{k}_1)+n_2\mathbf{k}_2+(\mathbf{k}_2\text{-}\mathbf{k}_1)\right].
\end{aligned} \tag{7}$$

The situation optimal for the considered effect is that of two electromagnetic waves that form a standing electromagnetic wave, and $\mathbf{k}_3 \perp \mathbf{A}_1$. It follows then, since the electromagnetic field is transverse, that the arguments of the Bessel functions vanish, so that a nonzero contribution is made by the Bessel functions with $n_1 = n_2 = 0$. The S-matrix element takes the form

$$S_{fi} = -(2\pi)^4 i \frac{e\mathbf{A}_3\mathbf{p}}{2mcV} \frac{e^2 A_1^2 (\mathbf{k}_2 - \mathbf{k}_1)^2}{(2mc^2)^2 \hbar (\mathbf{v}(\mathbf{k}_2 - \mathbf{k}_1))^2} \delta\left(\frac{E'-E}{\hbar} \pm \omega_3\right) \delta\left[\frac{(\mathbf{p}-\mathbf{p}')}{\hbar} \mp \mathbf{k}_3 + \pm 2\mathbf{k}_1)\right]. \tag{8}$$

From the energy and momentum conservation laws inherent in the delta functions of (8), we obtain an expression for the test-wave frequency at which amplification is possible

$$\omega_3 = 2(v/c)\omega_1 \cos\theta, \tag{9}$$

where $\theta$ is the angle between the vectors $\mathbf{p}$ and $\mathbf{k}_1$ ($0 \leq \theta \leq \pi/2$). Knowing the S-matrix element we can obtain the total probabilities of the induced emission and absorption processes. The total probabilities must be averaged over the initial energy distribution $f(E)$ of the electrons in the beam. We assume that the function $f(E)$ is normalized to unity, i.e., $\int f(E)dE = 1$, and that the function $f(E)$ has a width $\Delta E \ll E$.

We write then the expression for the emission and absorption differential probabilities in the form

$$dW_{e,a} = \left|S_{if}^{(e,a)}\right|^2 \frac{d\mathbf{p}'}{(2\pi)^3} V f(E) dE. \tag{10}$$

Substituting (8) in (10) and integrating the result with respect to $d\mathbf{p}'$, we get

$$W_e^{(1)} = 2\pi t \left(\frac{e\mathbf{A}_3\mathbf{p}}{2mc}\right)^2 \left[\frac{e^2 A_1^2}{(2mc^2)^2}\right]^2 \left(\frac{c}{v\cos\theta}\right)^4 \int \delta\left(\frac{E'-E}{\hbar} + \omega_3\right) f(E) dE,$$

$$W_a^{(1)} = 2\pi t \left(\frac{e\mathbf{A}_3\mathbf{p}}{2mc}\right)^2 \left[\frac{e^2 A_1^2}{(2mc^2)^2}\right]^2 \left(\frac{c}{v\cos\theta}\right)^4 \int \delta\left(\frac{E'-E}{\hbar} - \omega_3\right) f(E) dE. \tag{11}$$

We assume next that $\mathbf{p}$ is directed along $\mathbf{A}_3$. The $\delta$ functions in (11) can be represented in the form

$$\delta(E' - E \pm \omega_3) = \frac{\delta(E - E_{e,a})}{|\partial E'/\partial E - 1|} = \frac{\delta(E - E_{e,a})pc}{2\hbar\omega_1 \cos\theta}, \tag{12}$$

where $E_{e,a} = E_0 \pm \delta E$ is the energy of the electrons that emit or absorb a photon of a given frequency $\omega_3$:

$$E_0 = \frac{mc^2}{8\cos^2\theta}\left(\frac{\omega_3}{\omega_1}\right)^2, \quad \delta E = \frac{\hbar\omega_3}{2\cos^2\theta}, \quad \delta E \ll \Delta E, \tag{13}$$

where $\Delta E$ is the width of the distribution function.

Integrating with respect to dE in (11) and using (12), we obtain for the difference between the total probabilities for emission and absorption of a photon of frequency $\omega_3$:

$$\Delta W^{(1)} = 4\pi t \frac{(eA_3)^2 E^2}{mc^2\omega_3}\delta E\frac{df}{dE}\left[\frac{e^2 A_1^2}{(2mc^2)^2}\right]^2\left(\frac{c}{v\cos\theta}\right)^4. \tag{14}$$

In the derivation of (14) we used the approximate equality

$$f(E_e) - f(E_a) = 2\delta E\, df/dE, \tag{15}$$

the derivative $df/dE$ is taken at the point $E = E_0$.

The gain of the electromagnetic wave is determined in the linear regime by the relation

$$G^{(1)} = 8\pi(j/e)\Delta W^{(1)}\hbar\omega_3/c\varepsilon_3^2; \tag{16}$$

where $\varepsilon_3$ is the amplitude of the electromagnetic field intensity of the amplified wave and $j$ is the electron-current density.

**THE GAIN**

Substituting in (16) the expression for $\Delta W^{(1)}$ from (14), we get

$$G^{(1)} = 16\pi^2(j/e)r_0 t\lambdabar_3 E^2\left[\frac{e^2 A_1^2}{(2mc^2)^2}\right]^2\left(\frac{c}{v\cos\theta}\right)^4\frac{1}{\cos_2\theta}\frac{df}{dE}, \tag{17}$$

where $r_0$ is the classical electron radius. Relation (17) was obtained by using (13). We make one remark. Expression (17) for the gain is valid if $\Delta E/E > 1/\omega_1 t$ (t is the interaction time). Further estimates show that this criterion is valid. In the case of the inverse condition $\Delta E/E < 1/\omega_1 t$, however, the spontaneous-emission line is determined by the homogeneous width connected with the finite region of the interaction between the electrons and the field. The formal transition from $df/dE$ to homogeneous broadening is analogous to that in [52,53].

If the distribution function width is $\Delta E \gg \hbar\omega_3 / 2\cos^2\theta$, we have $df/dE \approx 1/(\Delta E)^2$. Taking this fact into account, the expression for the gain can be written in the form

$$G^{(1)} = 16\pi^2 (j/e) r_0 t \lambdabar_3 \left(\frac{E}{\Delta E}\right)^2 \left(\frac{e^2 \varepsilon_1 \lambdabar_1}{2mc^2}\right)^4 \left(\frac{c}{v}\right)^4 \frac{1}{\cos_2\theta}, \tag{18}$$

where $\varepsilon_1$ is the amplitude of the electric field strength.

The expression for the gain was obtained subject to validity of relations (20) of [58]. These relations are violated in the classical limit $\hbar\omega_1 \to 0$.

**CONCLUSION**

The foregoing result can be interpreted as follows. Nonrelativisitc electrons interacting with a standing electromagnetic wave can absorb photons from one wave and transfer them to another. This rescattering changes the electron energy by an amount of the order of the recoil energy $2\hbar\mathbf{k}\cdot\mathbf{v}$ and the energy of the quantum emitted in the recoil process can be of the same order.

We present a numerical example. We estimate the gain in the case of an amplified wavelength $\lambdabar_3 = 50\,\mu m$. We choose the electron-beam parameters to be: $j = 30 A/cm^2$, $\Delta E/E = 10^{-3}$, $v/c = 0.66$, and d (the beam diameter) = 1 mm.

Let the standing pump wave be produced by a $CO_2$ laser with the following parameters: $\lambda^{(1)} = 10.6\,\mu m$, $E_0$ (energy in the pulse) = 1 kJ, and pulse duration $\Delta t = 10 ns$. In this case $\cos\theta = 0.15$ and the gain $G^{(1)}$ calculated from Eq. (18) is found to be unity.

It can be easily verified that the aid of the above parameters that in our case $\Delta E/E < 1/\omega_1 t$, therefore the decisive factor is the inhomogeneous broadening which takes into account the initial energy distribution of the electrons in the beam.